\documentclass[aps,prb,twocolumn,showpacs,floatfix]{revtex4}

\usepackage{amsmath,graphicx}

\begin{document}
\preprint{Submitted to Phys. Rev. B}


\title{Electronic states and cyclotron resonance in
$p$-type InMnAs and InMnAs/(Al,Ga)Sb at ultrahigh magnetic fields}

\author{Y. Sun}
\author{F. V. Kyrychenko}
\author{G. D. Sanders}
\author{C. J. Stanton}
\affiliation{Department of Physics, University of Florida, Box 118440\\
Gainesville, Florida 32611-8440}

\author{G. A. Khodaparast}
\thanks{Present address: Physics Department, Virginia Tech,
Blacksburg, VA.}
\author{J. Kono}
\affiliation{Department of Electrical and Computer Engineering, Rice University, Houston, Texas
77005}

\author{Y. H. Matsuda}
\affiliation{Department of Physics, Faculty of Science, Okayama University, 3-1-1 Tsushimanaka,
Okayama 700-8530, Japan}

\author{H. Munekata}
\affiliation{Image Science and Engineering Laboratory, Tokyo Institute of Technology, Yokohama, Kanagawa
226-8503}

\date{\today}


\begin{abstract}

We present a theoretical and experimental study on electronic and magneto-optical
properties of p-type paramagnetic InMnAs dilute magnetic semiconductor alloys and
ferromagnetic p-type InMnAs/(Al,Ga)Sb thin films in ultrahigh ($>$ 100 T)
external magnetic fields \textbf{B}. We use an 8 band Pidgeon-Brown
model generalized to include the wavevector dependence of the electronic
states along \textbf{B} as well as $s$-$d$ and $p$-$d$ exchange interactions
with localized Mn $d$-electrons. In paramagnetic p-InMnAs alloys, we compute the
spin-dependent electronic structure as a function of Mn doping and examine
how the valence band structure depends on parameters such as the {\it sp-d}
exchange interaction strength and effective masses. The cyclotron resonance
(CR) and magneto-optical properties of InMnAs are computed using Fermi's golden
rule. In addition to finding strong CR for hole-active polarization
in p-type InMnAs, we also find strong CR for electron-active polarization.
The electron-active CR in the valence bands results from
transitions between light and heavy hole Landau levels and is seen in
experiments. In ferromagnetic p-InMnAs/(Al,Ga)Sb, two strong CR peaks are
observed which shift with position and increase in strength as the Curie
temperature is approached from above. This transition takes place well above the
Curie temperature and can be attributed to the increase in magnetic ordering
at low temperatures.

\end{abstract}


\maketitle

\section{Introduction}
\label{Introduction}

Electronic and optical properties of In$_{1-x}$Mn$_{x}$As
dilute magnetic semiconductors (DMS) are important
for designing ferromagnetic heterostructures, which
may in turn prove useful in the fabrication of spintronic
devices \cite{Munekata92.342,Ohno92.2664,Ohno96.363,Matsukura98.R2037,
Kikkawa98.139,Ohno99.790}. Since InMnAs is a narrow gap semiconductor, it
can serve as a prototype for studying electronic, spin, and ferromagnetic
properties of dilute magnetic semiconductor alloys. The results for InMnAs
can easily be generalized to the other dilute magnetic semiconductors.
Up to now, ferromagnetism in III-V ferromagnetic semiconductors has only been
observed in $p$-doped samples and it is believed that the ferromagnetic
exchange is mediated by itinerant holes \cite{Sanvito01.165206,Konig}.
Ferromagnetism strongly depends on the hole density and has been
observed in $p$-doped InMnAs samples having hole densities greater than
$10^{19}\mbox{ cm}^{-3}$. The nature of these itinerant holes is still an
open question. Therefore, studying the properties of the valence bands in
InMnAs is of fundamental importance in fully understanding the exchange
mechanism in dilute magnetic semiconductors.

Cyclotron resonance (CR) is a powerful method for
determining band parameters, especially in a DMS alloy,
whose bandstructure is very sensitive to external magnetic fields.
However, owing to strong disorder and scattering by Mn impurities in InMnAs,
one needs to go to ultrahigh magnetic fields (50-100 Tesla) where
$\omega_c\tau > 1 $ in order to observe CR.  Megagauss CR
studies have been performed on $n$-type CdMnTe \cite{Matsuda02.115202} and
$n$- and $p$-type InMnAs
\cite{Matsuda01.219,Zudov02.161307,Khodaparast02.320,Khodaparast03.107,Sanders03.6897,Sanders03.165205}
and have shown that the Mn impurities have a significant
effect on the electronic structure of DMS alloys.

In order to study the electronic and optical properties of DMS alloys,
we utilize a modified 8 band Pidgeon-Brown effective mass
Hamiltonian \cite{Pidgeon66.575} which is generalized to include the dependence
of the bandstructure on the wavevector $k$ parallel to \textbf{B} as well as
the $sp-d$ exchange interaction between the itinerant carriers and localized Mn spins.

This paper is organized as follows. In section \ref{theory section},
the theory used in this paper will be discussed in detail including the effective
mass Hamiltonian and the magneto-optical absorption. In section \ref{results}, we
will discuss our results and investigate cyclotron resonance absorption in bulk
paramagnetic DMS samples and ferromagnetic DMS thin films.
Finally, our conclusions will be given in section \ref{conclusions}. In the
appendix, we discuss the validity of the 8 band $\textbf{k} \cdot \textbf{p}$
theory for magnetic fields up to 100 T.

\section{Theory}
\label{theory section}

In this section, we describe the theoretical model we use to
analyze Megagauss CR data in bulk dilute magnetic semiconductor alloys.
Our effective mass treatment of InMnAs was described in
Ref.~\onlinecite{Sanders03.6897} where {\it n}-type dilute magnetic
semiconductors were considered. Cyclotron resonance in {\it p}-type
dilute magnetic semiconductors is more complicated owing to the
greater complexity of the valence bands. The formalism however
follows our previous work on $n$-type systems and the reader is referred
to this reference for details of the effective mass formalism. Here
we simply outline the important points.

The effective mass Hamiltonian for In$_{x}$Mn$_{1-x}$As in a
magnetic field directed along the $z$ axis can be expressed as the
sum of three Hamiltonians which account for the Landau, Zeeman,
and $sp-d$ exchange interactions between $s$ and $p$ electrons and
localized Mn $d$ electrons. Thus,
\begin{equation}
H = H_L + H_Z + H_{Mn}. \label{HTotal}
\end{equation}
where explicit expressions for effective mass Hamiltonians,
$H_L$, $H_Z$, and $H_{Mn}$ are given in Ref.~\onlinecite{Sanders03.165205}.

Following Pidgeon and Brown, we find it convenient to separate the
eight Bloch basis states into an upper set and a lower set which decouple
at the zone center, i.e. $k_{z}=0$. The Bloch basis set we choose
for the upper set are
\begin{subequations}
\label{upperset}
\begin{eqnarray}
\arrowvert 1 \rangle &=& \arrowvert \frac{1}{2}, +\frac{1}{2} \
\rangle= \arrowvert S \uparrow \rangle
\\
\arrowvert 2 \rangle &=& \arrowvert \frac{3}{2}, +\frac{3}{2} \
\rangle= \frac{1}{\sqrt{2}} \arrowvert (X + i Y) \uparrow \rangle
\\
\arrowvert 3 \rangle &=& \arrowvert \frac{3}{2}, -\frac{1}{2}
\rangle= \frac{1}{\sqrt{6}} \arrowvert (X - i Y) \uparrow +2 Z
\downarrow \rangle
\\
\arrowvert 4 \rangle &=& \arrowvert \frac{1}{2}, -\frac{1}{2}
\rangle= \frac{i}{\sqrt{3}} \arrowvert -(X - i Y) \uparrow +Z
\downarrow \rangle
\end{eqnarray}
\end{subequations}
which correspond to electron spin up, heavy hole spin up, light
hole spin down, and split off hole spin down. Likewise, the Bloch
basis states for the lower set are
\begin{subequations}
\label{lowerset}
\begin{eqnarray}
\arrowvert 5 \rangle &=& \arrowvert \frac{1}{2},-\frac{1}{2}
\rangle= \arrowvert S \downarrow \rangle
\\
\arrowvert 6 \rangle &=& \arrowvert \frac{3}{2}, -\frac{3}{2} \
\rangle= \frac{i}{\sqrt{2}} \arrowvert (X - i Y) \downarrow
\rangle
\\
\arrowvert 7 \rangle &=& \arrowvert \frac{3}{2}, +\frac{1}{2}
\rangle= \frac{i}{\sqrt{6}} \arrowvert (X + i Y) \downarrow -2 Z
\uparrow \rangle
\\
\arrowvert 8 \rangle &=& \arrowvert \frac{1}{2}, +\frac{1}{2}
\rangle= \frac{1}{\sqrt{3}} \arrowvert (X + i Y) \downarrow +Z
\uparrow \rangle
\end{eqnarray}
\end{subequations}
corresponding to electron spin down, heavy hole spin down, light
hole spin up, and split off hole spin up.

For the vector potential,
we choose the Landau gauge
\begin{equation}
\vec{A} = -B \ y \ \hat{x} \label{Gauge}
\end{equation}
from which we obtain $\vec{B} = \vec{\nabla} \times \vec{A} = B \
\hat{z}$.
With the choice of Landau Gauge (\ref{Gauge}), translational symmetry in
the $x$ direction is broken while translational symmetry along the
$y$ and $z$ directions is maintained. Thus $k_y$ and $k_z$ are
good quantum numbers and the envelope of the effective mass
Hamiltonian (\ref{HTotal}) can be written as
\begin{equation}
{\cal{F}}_{n,\nu} = \frac{e^{i(k_y y + k_z z)}}{\sqrt{{\cal{A}}}}
\left[
\begin{array}{l}
a_{1,\nu} \ \phi_{n-1}   \\
a_{2,\nu} \ \phi_{n-2} \\
a_{3,\nu} \ \phi_{n} \\
a_{4,\nu} \ \phi_{n} \\
a_{5,\nu} \ \phi_{n} \\
a_{6,\nu} \ \phi_{n+1} \\
a_{7,\nu} \ \phi_{n-1}   \\
a_{8,\nu} \ \phi_{n-1}
\end{array}
\right] \label{Fn}
\end{equation}

In Eq. (\ref{Fn}), $n$ is the Landau quantum number associated
with the Hamiltonian matrix, $\nu$ labels the eigenvectors
in order of increasing energy,
${\cal{A}} = L_x L_y$ is the cross sectional area of the sample in
the $xy$ plane, $a_{i,\nu}(k_z)$ are complex expansion coefficients for the
$\nu$-th eigenstate which depend explicitly on $k_z$, and $\phi_n(\xi)$
are harmonic oscillator eigenfunctions evaluated at
$\xi = x - \lambda^2 k_y$. The magnetic length, $\lambda$, is
\begin{equation}
\lambda = \sqrt{\frac{\hbar c}{e B}}
        = \sqrt{\frac{\hbar^2}{2 m_0} \ \frac{1}{\mu_B B}}.
\label{lambda}
\end{equation}
where $\mu_B=5.789 \times 10^{-5}\ \mbox{eV/Tesla}$ is the Bohr
magneton and $m_0$ is the free electron mass.

Substituting ${\cal{F}}_{n,\nu}$ from Eq. (\ref{Fn}) into the
effective mass Schr\"{o}dinger equation with $H$ given by Eq.
(\ref{HTotal}), we obtain a matrix eigenvalue equation
\begin{equation}
H_n \ F_{n,\nu} = E_{n,\nu}(k_z) \ F_{n,\nu},
\label{Schrodinger}
\end{equation}
that can be solved for each allowed value of the Landau quantum
number, $n$, to obtain the Landau levels $E_{n,\nu}(k_z)$. The
components of normalized eigenvectors, $F_{n,\nu}$, are the
expansion coefficients, $a_i$.
Since the harmonic oscillator functions, $\phi_{n'}(\xi)$, are
only defined for $n' \ge 0$, it follows from Eq. (\ref{Fn}) that
$F_{n,\nu}$ is defined for $n \ge -1$. In solving the
effective mass Schr\"{o}dinger equation (\ref{Schrodinger}),
rows and columns for which $n' < 0$ are deleted in the effective
mass Hamiltonian. The energy levels are
denoted $E_{n,\nu}(k_z)$ where $n$ labels the Landau level and
$\nu$ labels the eigenenergies belonging to the same Landau level
in ascending order.

\begin{figure} [tbp]
\includegraphics[scale=0.35]{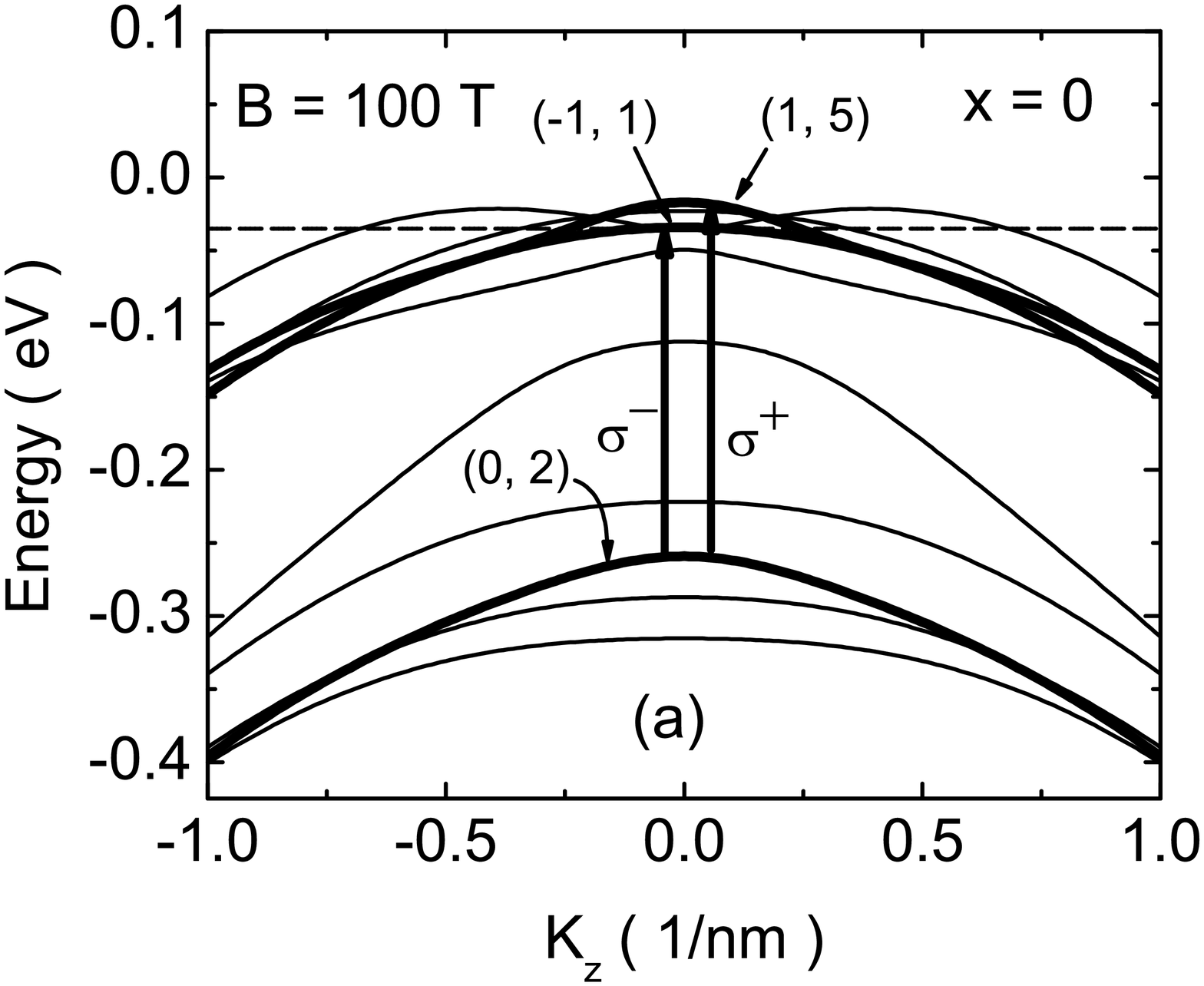}
\includegraphics[scale=0.35]{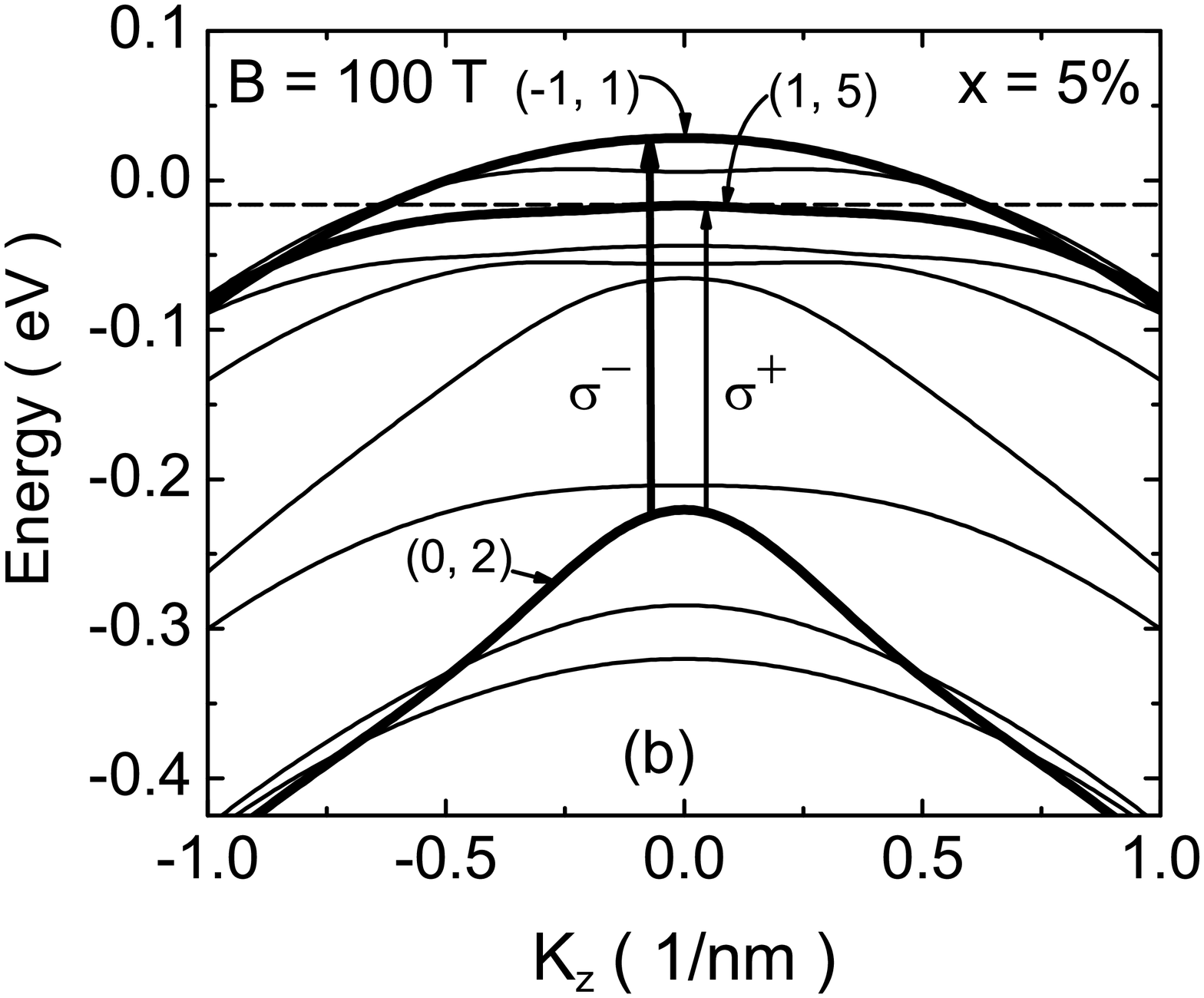}
\caption{Valence band structure for $T = 30\ \mbox{K}$ and $B = 100\ \mbox{T}$
for In$_{1-x}$Mn$_x$As alloys having (a) $x=0$\% and (b) $x=5$\%. For $x=0$\%,
the first HH state,$H_{-1,1}$, lies below the light hole state $L_{1,5}$. For $x=5\%$,
the order of these two states is reversed. Two possible CR transitions are
shown using upward arrows, namely an h-active ($\sigma^{-}$) transition between
$H_{0,2}$ and $H_{-1,1}$, and an e-active ($\sigma^{+}$) transition from
$H_{0,2}$ to $LH_{1,5}$. The dashed lines are the Fermi energies for a hole
density of $10^{19} \mbox{cm}^{-3}$.} \label{VBx}
\end{figure}

The magneto-optical absorption coefficient at the photon energy
$\hbar \omega$ is \cite{Bassani}
\begin{equation}
\alpha(\hbar \omega)= \frac{\hbar \omega}{(\hbar c) n_r}\
\epsilon_2(\hbar \omega)
\end{equation}
where $\epsilon_2(\hbar \omega)$ is the imaginary part of the
dielectric function and $n_r$ is the index of refraction. The
imaginary part of the dielectric function is found using Fermi's
golden rule as described in Ref.~\onlinecite{Sanders03.165205} and
the material parameters we use are listed in Table I of
Ref.~\onlinecite{Sanders03.165205}.

\section {Results and Discussion}
\label{results}

In$_{1-x}$Mn$_x$As was the first III-V DMS material in which ferromagnetism was
observed \cite{Ohno92.2664,Munekata93.2929}. While Curie
temperatures can be above 50 K, in many instances, the Curie
temperature in InMnAs is below 5 K. In our experiments the samples can be
considered as paramagnetic and we consider this limit in what follows
using the formalism described in Ref.~\onlinecite{Sanders03.165205}.

\subsection {Valence subband structure}
\label{valence subband structure}

Solving the effective mass Schr\"{o}dinger equation (\ref{Schrodinger}),
we obtain the valence subband structure as a function of Mn
concentration, $x$ and wavevector, $k_{z}$. The valence subband
structure in In$_{1-x}$Mn$_{x}$As for $T = 30$ K and $B = 100$ Tesla is
shown in Fig.~\ref{VBx} for $x=0$ and $x=5$\%.
Comparing Fig.~\ref{VBx} (a) and (b), we see that doping with Mn leads to
large splitting of the valence bands due to the exchange interaction between itinerant holes and
localized Mn magnetic moments. In this subband structure, the Landau level with
$n=-1$ (obtained by diagonalizing a $1 \times 1$ effective mass Hamiltonian)
is a pure heavy hole (HH) spin down state with a simple parabolic
dispersion relation, which we label $H_{-1,1}$.

As seen in Fig.~\ref{VBx} (b), the most pronounced effect of Mn doping
on the computed valence band Landau levels is the reversal in energy of two
states, namely the pure heavy hole $H_{-1,1}$ state and the $L_{1,5}$ state which
is primarily a light hole spin up state near the zone center.
These two states are the lowest-lying hole levels in undoped
and doped samples, respectively, so in the limit of low carrier density,
they will be the only occupied states.
Thus, a major effect of Mn doping is to change the spin
state of the system through the exchange interaction between the itinerant
holes and the localized Mn spins.

\subsection {Cyclotron resonance absorption}

\begin{figure} [tbp]
\includegraphics[scale=.65]{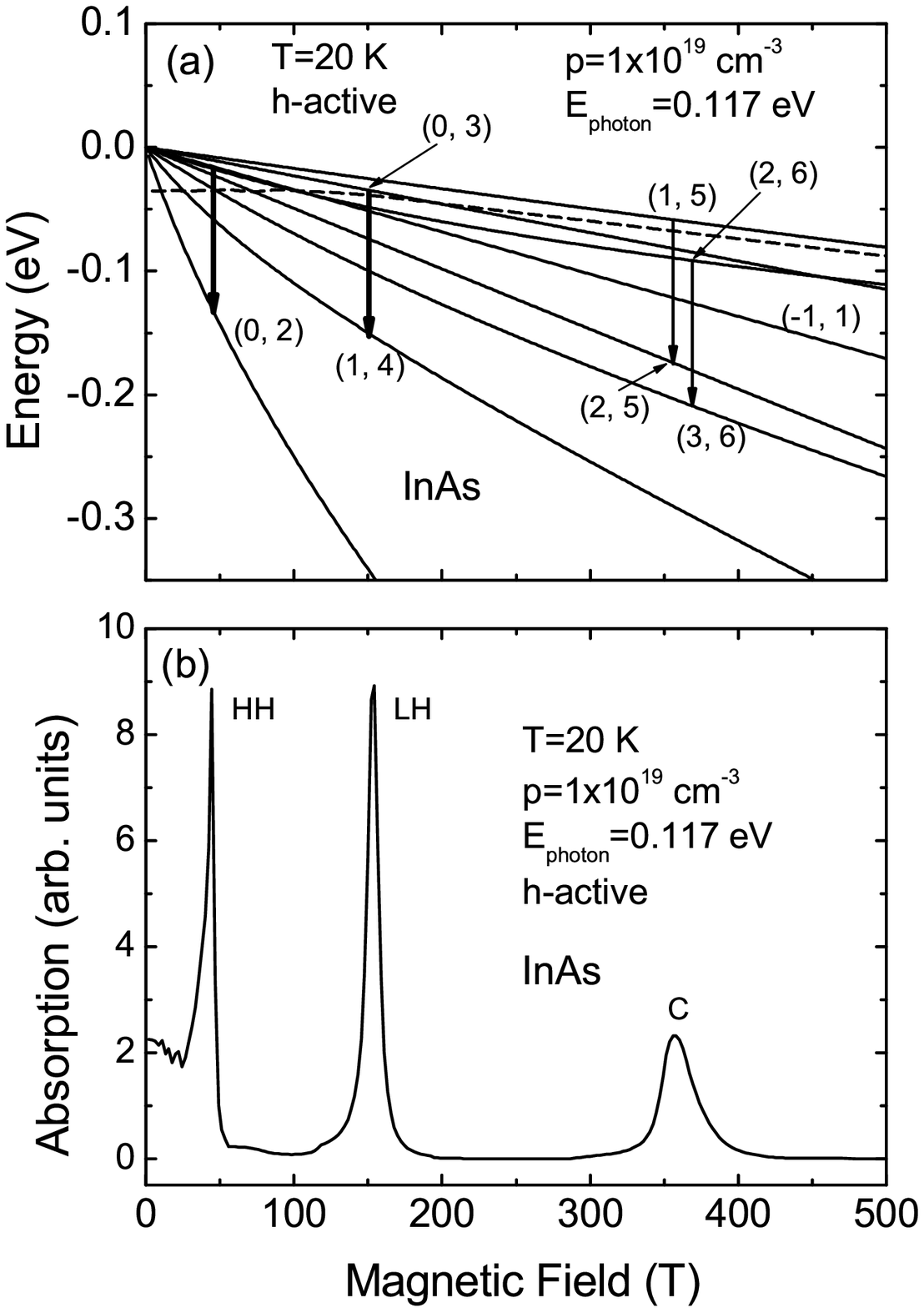}
\includegraphics[scale=0.3750]{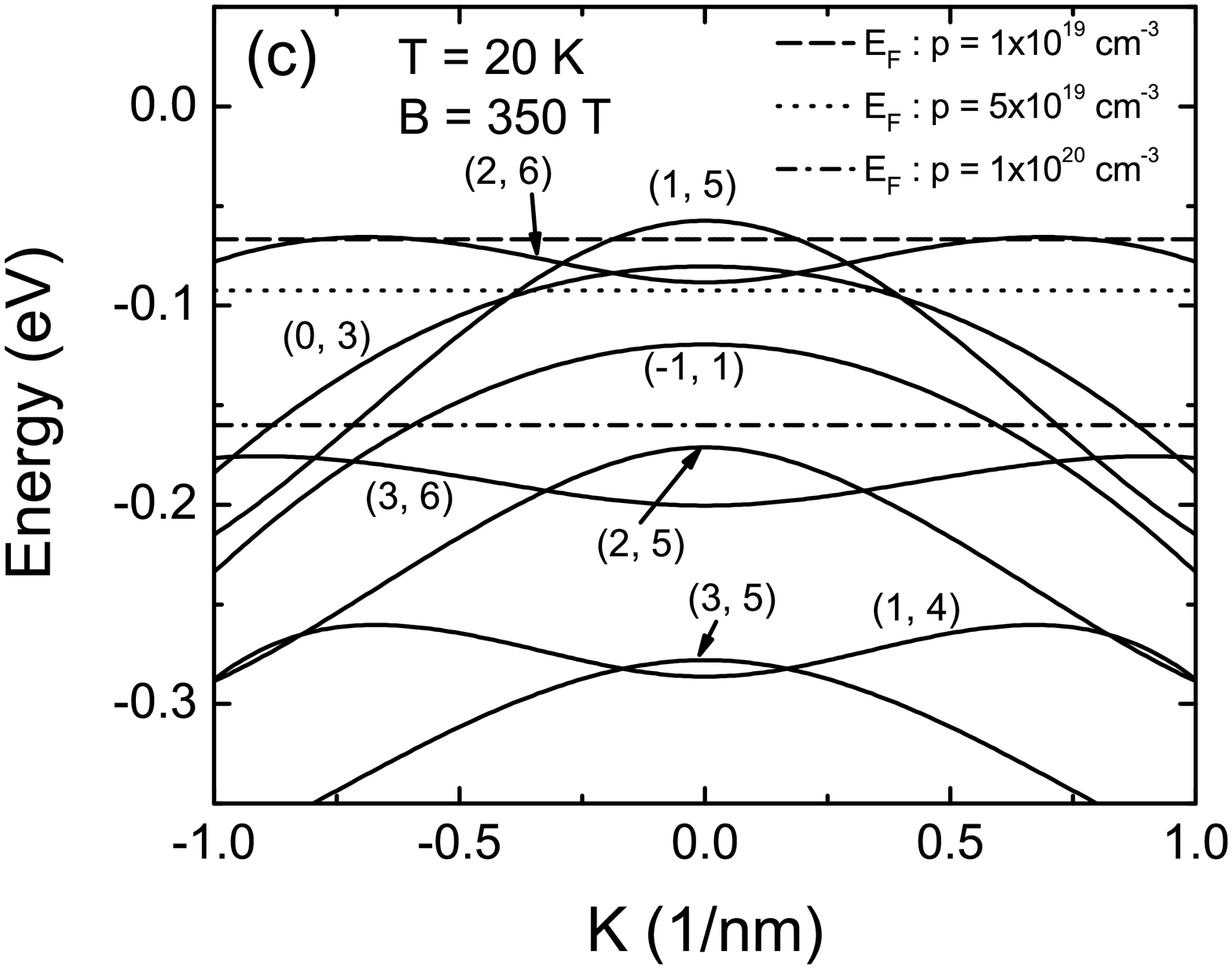}
\caption{
The upper panel (a) shows the $k=0$ valence band Landau levels as a
function of B and the Fermi level for $p=10^{19} \ \mbox{cm}^{-3}$
(dashed line). The CR absorption in $p$-type InAs is shown in (b)
for h-actively polarized radiation with
$\hbar\omega=0.117\ \mbox{eV}$  at $T=20\ \mbox{K}$ and
$p=10^{19} \ \mbox{cm}^{-3}$. A FWHM linewidth of 4 meV is assumed.
The k-dependent Landau subband structure at B = 350 Tesla. is shown in (c).}
\label{VBb}
\end{figure}

Our theoretically computed CR absorption in undoped InAs
for h-active circularly polarized radiation with photon energy
$\hbar\omega=0.117\ \mbox{eV}$  at $T = 20\ \mbox{K}$ is shown in
Fig.~\ref{VBb}(b) for a hole concentration of
$p = 10^{19} \ \mbox{cm}^{-3}$. The CR absorption spectrum is
broadened with a FWHM linewidth of $4$ meV which is narrower than in
the experimental situation.

To help us understand the origin of the CR absorption peaks,
the Landau levels at $k = 0$ are shown as a function of the applied
magnetic field in the fan diagram in Fig.~\ref{VBb} (a).
The Fermi level for $p=10^{19}\mbox{cm}^{-3}$ is indicated by the dashed
line in the fan diagram. At a hole density of $10^{19}\mbox{cm}^{-3}$
only Landau levels very close to the valence band edge are occupied by holes.
The transitions responsible for the strong CR absorption peaks are
indicated by the vertical arrows. Holes excited from the low-lying
Landau levels give rise to three CR peaks. The CR absorption peak near
40 Tesla is due to a transition between the Landau subbands $H_{-1,1}$
and $H_{0,2}$. Near the zone center, the $H_{0,2}$ level is primarily
HH spin down which accounts for our use of the $'H'$ designation for
this subband. The CR peak near 150 Tesla is due to a transition between
the $L_{0,3}$ and $L_{1,4}$ states where $'L'$ indicates the states
are primarily LH in character at $k=0$.

There is another CR absorption peak around 350 Tesla, labeled 'C', which can be
resolved into a closely spaced pair of HH and LH peaks indicated by the
two vertical arrows in the fan diagram in Fig.~\ref{VBb} (a), i.e. a light
hole $L_{1,5}$ to $L_{2,5}$ transition and a heavy hole $H_{2,6}$ to $H_{3,6}$
transition. The $H_{2,6}$ to $H_{3,6}$ transition demands some explanation since
it would appear from the fan diagram that both initial and final states are
below the Fermi level and contain no holes. So how can there be a CR
absorption peak for this transition? The answer can best be seen by
plotting the k-dependent Landau level subband structure at 350 Tesla.
This is done in Fig.~\ref{VBb} (c) where Fermi levels are
shown for hole concentrations of
$p=10^{19}\ \mbox{cm}^{-3}$, $5 \times 10^{19}\ \mbox{cm}^{-3}$, and
$10^{20}\ \mbox{cm}^{-3}$.
The $H_{2,6}$ Landau subband exhibits a camel back structure and
at a hole concentration of $p=10^{19}\ \mbox{cm}^{-3}$
only states near $k=0.6 \ \mbox{nm}^{-1}$ are occupied while states at
the $\Gamma$ point are empty. It is clear that the $H_{2,6}$ to
$H_{3,6}$ transition takes place between states near
$k=0.6 \ \mbox{nm}^{-1}$. As an aside, our analysis reveals that the
transition at the $\Gamma$ point between the two states is actually
forbidden while mixing of states away from the $\Gamma$ point gives
rise to nonvanishing optical matrix elements.

Finally, we should note that for $B < 30$ Tesla, higher order
Landau levels become occupied and excitations of holes from these
subbands are responsible for the downward sloping plateau seen in
the cyclotron resonance absorption in Fig.~\ref{VBb} (b)
for $B < 30 \ \mbox{T}$.
%
\begin{figure}[tbp]
\includegraphics[scale=0.375]{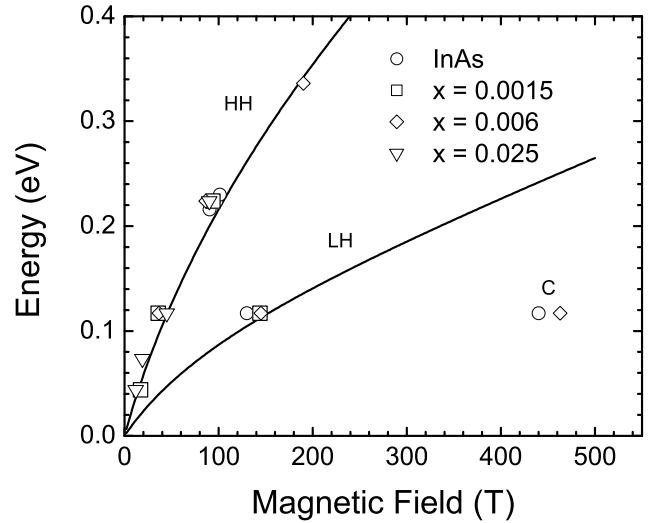}
\caption{Observed hole-active cyclotron resonance peak positions
for the HH, LH and 'C' features for four
samples with different Mn concentrations. The two solid curves are
calculated peak positions based on the 8-band effective mass model
described in the text.}
\label{HLB}
\end{figure}

We find that for the low Mn concentrations encountered in experiments,
the CR peak positions in In$_{1-x}$Mn$_x$As are insensitive to $x$.
In Fig.~\ref{HLB}, we plot the lowest lying
HH, LH  and 'C' CR peak positions as functions of magnetic field for
$\hbar \omega = 0.117 \ \mbox{eV}$ in four samples having
Mn concentrations from $x=0$ to $x=0.025$. The solid curves
are the theoretically computed CR peak positions for the
HH and LH CR peaks for InAs in the paramagnetic limit. Thus, doping with
low Mn concentrations does not alter the band structure significantly.

\begin{figure} [tbp]
\includegraphics[scale=0.475]{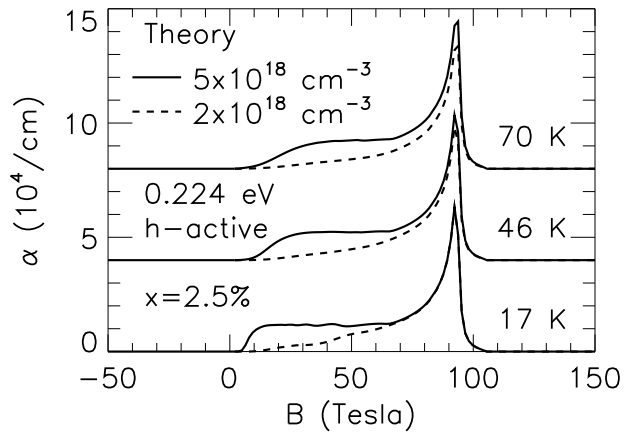}
\caption{CR dependence on hole densities. Due to the Fermi filling effect, the
CR spectrum with higher hole density has a plateau at low fields, which
becomes smoother with increasing temperature.}
\label{CRdensity}
\end{figure}

While the CR peak positions may be insensitive to
the Mn concentration, the CR spectra depend sensitively
on the hole density due to Fermi filling effects.
Fig.~\ref{CRdensity} shows theoretical h-active CR absorption
spectra in In$_{0.975}$Mn$_{0.025}$As for photon energies having
$\hbar \omega = 0.224 \ \mbox{eV}$. We plot the CR absorption
at two different hole concentrations,
$p=2 \times 10^{18} \ \mbox{cm}^{-3}$ and
$5 \times 10^{18} \ \mbox{cm}^{-3}$ and at three different
temperatures, namely $T = 17 \ \mbox{K}$, $46 \ \mbox{K}$
and $70 \ \mbox{K}$ assuming a narrow FWHM linewidth of 4 meV.

The CR absorption spectra exhibit asymmetric peaks with a broad
tail at low fields. Note that the width of the low field tail
depends on the free hole concentration. The curves with a hole density
of $5\times10^{18}\mbox{cm}^{-3}$ have a broad tail at
low fields, while the curves with a hole density of
$2\times10^{18}\mbox{cm}^{-3}$ have a narrow tail. The width of the low
field tail results from the population of higher order Landau
levels and the sharpness of the low field cutoff is attributed to the
sharpness of the Fermi distribution at low temperatures.
The HH to HH transition will take place even if the hole
density is low, because the $H_{-1,1}$ state is very close to the
valence band edge and is almost always occupied, partially or
fully. The higher order transitions need higher free hole densities to
have the corresponding levels occupied.

\begin{figure} [b]
\includegraphics[scale=0.7]{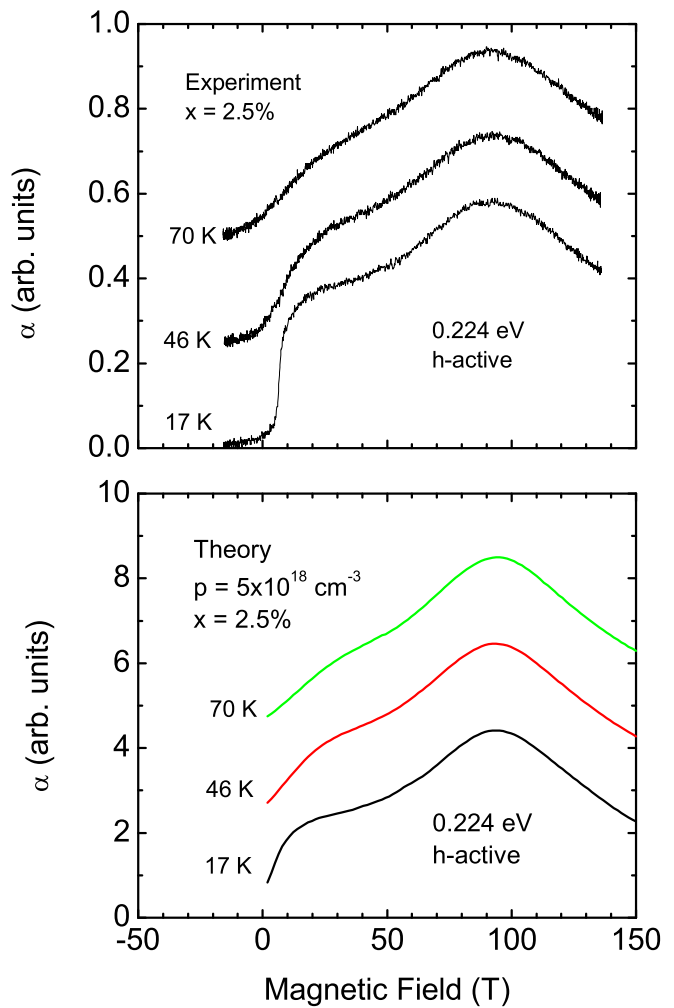}
\caption{Experimental CR and corresponding theoretical simulations. The low
temperature CR has an abrupt cutoff at low fields due to the fermi level
sharpening effect.}
\label{FLsharpening}
\end{figure}

Due to the nonparabolicity of the valence subbands, most CR spectra are
generally not symmetric. The asymmetry of the CR absorption curves in
Fig.~\ref{CRdensity} is seen very clearly.
Changes in temperature can also bring about changes in the spectral
lineshape due to Fermi level sharpening at low
temperatures. This can be seen in the experimental data shown in
Fig.~\ref{FLsharpening} (a) where the 17 K curve has an abrupt cutoff
at low magnetic fields. Fig.~\ref{FLsharpening} (b) shows the
results of our theoretical CR absorption spectra, in which the
broadening of the CR absorption curves is based on the experimental hole
mobilities.

\subsection {Sensitivity of cyclotron resonance energies to model parameters}

\begin{figure} [tbp]
\includegraphics[scale=0.6]{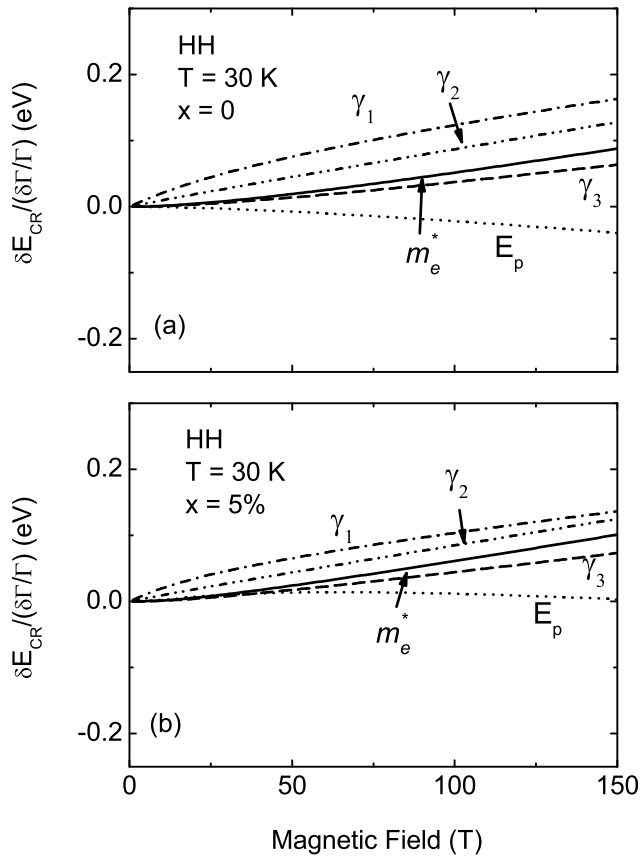}
\caption{Change in the heavy-hole cyclotron resonance energy
in In$_{1-x}$Mn$_x$As for a 10\% change in the model parameters
$\Gamma$ as a function of the resonance field at T = 30 K. The upper
panel (a) is for x = 0\% and the lower panel (b) is for x = 5\%.}
\label{PEOB}
\end{figure}

\begin{figure} [tbp]
\includegraphics[scale=0.6]{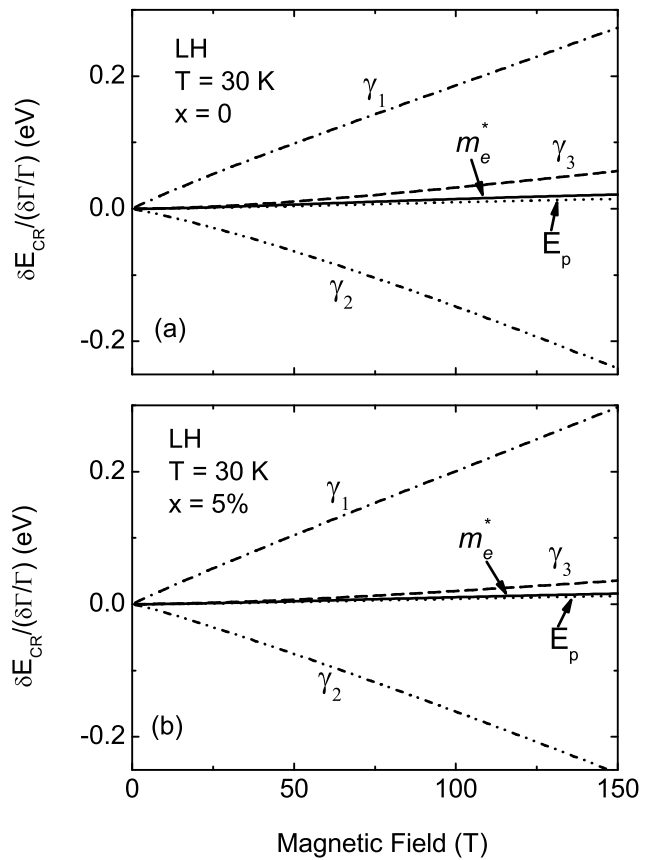}
\caption{Change in the light-hole cyclotron resonance energy
in In$_{1-x}$Mn$_x$As for a 10\% change in the model parameters
$\Gamma$ as a function of the resonance field at T = 30 K. The upper
panel (a) is for x = 0\% and the lower panel (b) is for x = 5\%.}
\label{PEOB2}
\end{figure}

The material parameters used in the effective mass Hamiltonian
can drastically change the valence band structure and the resulting CR
absorption spectra.
Figs.~\ref{PEOB} and \ref{PEOB2} shows the heavy-hole and light-hole
cyclotron resonance energy shift in In$_{1-x}$Mn$_x$As due to
a $10 \%$ change in a number of material parameters. We denote this shift
in the cyclotron resonance energy by $\delta E_{\textrm{CR}}/(\delta \Gamma/\Gamma)$.
The material parameters $\Gamma$ include the Luttinger parameters
$\gamma_{1}$, $\gamma_{2}$, $\gamma_{3}$, Kane's parameter
$E_{p}$, and the effective electron mass $m_e^*$. Shifts in the cyclotron
resonance photon energy are plotted as a function of the resonance field.
It should be remarked that the resonance photon energy is not constant but
varies with the resonance field. As expected, shifts in the cyclotron
resonance photon energy vanish if the resonance field vanishes.
In all cases the temperature is taken to be 30 K and results are
plotted for Mn concentrations of 0\% and 5\%.

Comparing Figs.~\ref{PEOB} (a) and \ref{PEOB2} (a) we find that
the LH transition in InAs is more sensitive to small changes in the material
parameters than the HH transitions. For instance, a $10\%$ change
in $\gamma_{1}$ will result in a $\sim 0.1 \ \mbox{eV}$ shift in
the LH cyclotron resonance energy, which in turn will result in
about a 100 T change in the cyclotron resonance field.
In Figs.~\ref{PEOB} (b) and \ref{PEOB2} (b) the sensitivity
of the CR energy to changes in the material parameters in
In$_{0.95}$Mn$_{0.05}$As are plotted. We find that 5\% Mn doping
doesn't strongly alter the sensitivity to small changes in the
material parameters.

\begin{figure} [tbp]
\includegraphics[scale=.45]{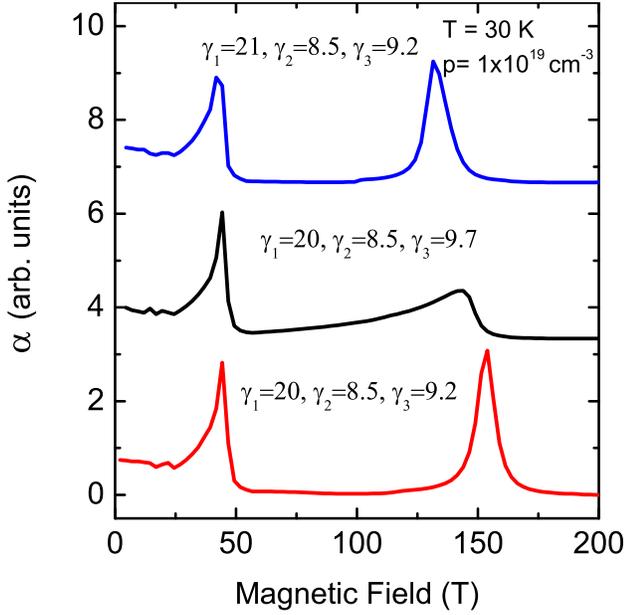}
\caption{Theoretical CR spectra of InAs using three different sets of
Luttinger parameters. The bottom curve is obtained using nominal values
of the Luttinger parameters. The light-hole transition is seen to be more
sensitive to small changes in the Luttinger parameters.}
\label{CRpara}
\end{figure}

It is instructive to demonstrate how the CR absorption
depends on the valence band Luttinger parameters. The nominal values
of the Luttinger parameters in InAs \cite{Vurgaftman01.5815} are
$\gamma_1=20$, $\gamma_2=8.5$ and $\gamma_3=9.2$ . In Fig.~\ref{CRpara}
we plot theoretical CR absorption spectra in InAs for three different
sets of Luttinger parameters while fixing all other material parameters
at their nominal values. The bottom curve in Fig.~\ref{CRpara} is based
on the nominal InAs Luttinger parameters. The upper curve illustrates
what happens when $\gamma_1$ is increased by 5\% and the middle
curve shows what happens when $\gamma_3$ is increased by 5\%.
In all three simulations, we assume that the temperature $T = 30 \mbox{K}$
and that the the hole concentration $p = 10^{19} \mbox{cm}^{-3}$.
>From Fig.~\ref{CRpara} it is clear that the HH CR resonance peak near
40 Tesla is relatively insensitive to small changes in the Luttinger
parameters while the position and shape of the LH CR feature near 150
Tesla is very sensitive to changes in the Luttinger parameters.
It can be thus be seen that the CR spectra depends sensitively on the
material parameters and may furnish an effective way to refine these
parameters.

\subsection {Selection rules, valence band mixing and e-active
CR in $p$-type semiconductors}

Selection rules for CR are a direct result of the conservation of
energy and angular momentum. For a free gas of electrons or holes,
CR can only be observed for certain circular polarizations.
For instance, in a free electron gas, CR transitions occur for
$\sigma^+$ circular polarization
while for a free hole gas, CR occurs only for
$\sigma^-$ circular polarization. As a result, CR in
$\sigma^+$ polarization is often called electron-active (e-active)
and $\sigma^-$ polarization is referred to as hole-active
(h-active).

The situation in real semiconductors, however, differs from that
of a gas of free positive or negative carriers. Recently,
we have experimentally observed e-active CR in $p$-doped InAs and
InMnAs. The temperature was quite low (12K) and the hole
concentration was high enough ($10^{19}\mbox{cm}^{-3}$) to safely
eliminate the possibility that the e-active CR comes from the
thermally excited electrons in the conduction band. We excluded
the existence of electrons in the interface or surface inversion
layers. Thus, the results suggest that e-active CR comes from the
valence band holes, in contradiction to the simple picture of a
free hole gas.

With a full treatment of the semiconductor energy band structure
within our modified Pidgeon-Brown model,
it is easy explain the e-active CR results in terms of the degeneracy
of the valence band states. Specifically,
having multiple valence band states
(heavy hole: $J =3/2, M_j=\pm 3/2$; light hole: $J=3/2, M_j= \pm 1/2$)
allows one to satisfy conservation of angular
momentum in cyclotron absorption for both $\sigma^+$ and
$\sigma^-$ polarizations.  Note that this is not a consequence of
valence band mixing between the heavy and light hole states
(though mixing between the conduction and valence bands is needed
to pick up oscillator strength for the transitions). It is simply
required to have degenerate valence band states with differing
$M_j$ quantum numbers.

The effective mass wavefunction is a product of the
rapidly oscillating periodic part of the Bloch basis functions
at the zone center $u_{\alpha}$ and the slowly varying envelope
function $\Phi_{\alpha}$,
\begin{equation}
  \Psi^i\equiv |i\rangle = \sum_{\alpha} \Phi_{\alpha}^i u_{\alpha},
\end{equation}
where the $\Phi$'s refer to the oscillator functions multiplied by
$k$ dependent coefficients, and the subscript $\alpha$'s refer to
the quantum numbers $n$ and $\nu$ in Eq. (\ref{Fn}). The summation
is performed over different energy bands. The optical matrix
element is given by
\begin{equation}\label{P}
  \langle f |\hat{\bf P}|i\rangle \approx \sum_{\alpha,\alpha'} \langle
  \Phi_{\alpha'}^f |\Phi_{\alpha}^i\rangle \langle u_{\alpha'}|\hat{\bf
  p}|u_{\alpha}\rangle+\langle u_{\alpha'}|u_{\alpha}\rangle \langle
  \Phi^f_{\alpha'}|\hat{\bf P}|\Phi^i_{\alpha}\rangle,
\end{equation}
where we used $\alpha$ and $\alpha'$ to refer to $n,\mu,k_{z}$
and $n',\mu',k_{z}$, respectively.

Two terms appear in Eq.~(\ref{P}). The second term, proportional
to the momentum matrix element between the envelope functions
describes optical transitions within the one-band model and
corresponds to that in the free electron gas. The first term,
being proportional to the momentum matrix element between periodic
parts of the Bloch functions, has interband nature and is present
only if conduction band - valence band mixing takes place. As
shown in Ref.~\onlinecite{Kacman71.629}, the first term dominates in both narrow
and wide gap semiconductors.

Substituting the wavefunctions in Eq. (\ref{Fn}) into Eq. (\ref{P}),
the selection rules for optical transitions in e-active and h-active
polarizations are obtained from
\begin{equation}\label{selection}
\langle \Psi_{n'}|\hat{P}_{\pm}|\Psi_n\rangle \propto \delta_{n',n
\pm 1}.
\end{equation}

We need to stress that in Eq.~(\ref{P}), both terms will
result in the same selection rules. This means that the CR
selection rules do not depend on the degree of conduction-valence
band mixing.
As a result of the selection rules, we see that $\sigma^+$
illumination leads to transitions which increase the manifold
quantum number by one ($n\rightarrow n+1$) while $\sigma^-$ leads
to transitions which decrease the manifold quantum number by one
($n\rightarrow n-1$).

In the conduction band, increasing the manifold quantum number
always increases the energy. As a result, only transitions with
increasing $n$ may take place in absorption, that is, only
e-active ($\sigma^+$) CR can be observed in the conduction band.

\begin{figure}[tbp]
\includegraphics[scale=0.5]{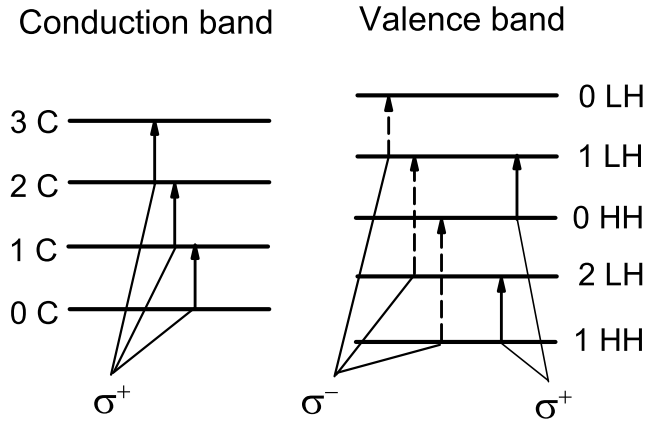}
\caption{Schematic diagram of Landau levels and CR transitions in
conduction and valence bands. Both h-active and e-active
transitions are allowed in the valence band because of the
degenerate valence band structure. Only e-active transitions are
allowed in the conduction band.}
\label{diag}
\end{figure}

The valence band, however, consists of two types of carriers:
heavy holes ($J =3/2, M_j=\pm 3/2$) and light holes
($J =3/2, M_j=\pm 1/2$). Each of them has their own Landau ladder in a
magnetic field. An increase of $n$ always implies a decrease in energy
\textit{but only within each ladder}. Like the conduction band case,
transitions within a ladder ($\rm HH\rightarrow HH$ or
$\rm LH\rightarrow LH$) can take place only in h-active ($\sigma^-$)
polarization. However, the relative position of the two ladders
can be such that interladder transitions ($\rm LH\rightarrow HH$)
in e-active polarization are allowed.
This process is schematically shown in Fig.~\ref{diag}. Note
that this figure is extremely simplified and should be used only
as a qualitative explanation of the effect.

\begin{figure} [tbp]
\includegraphics[scale=0.6]{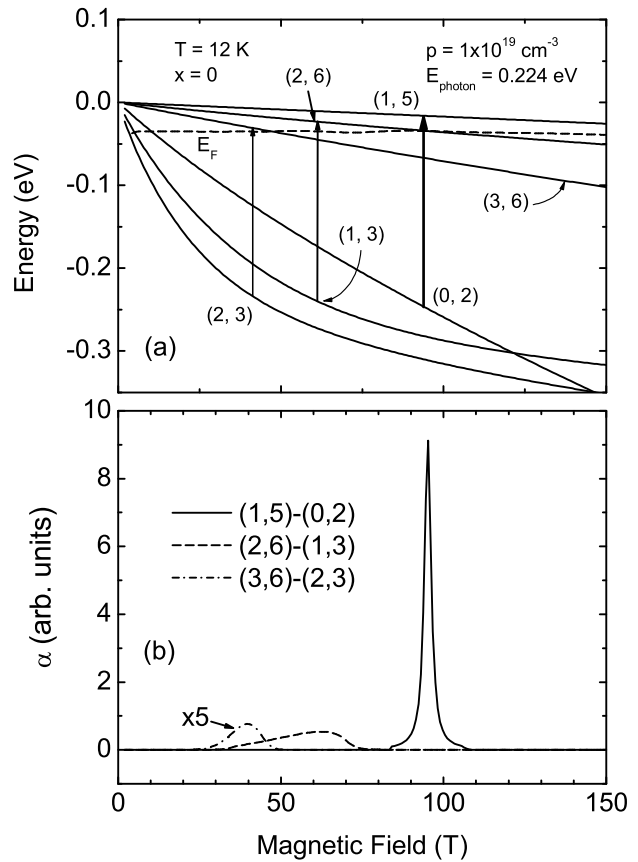}
\caption{The Landau level fan diagram (a) and the three strongest e-active
CR features at a photon energy of 0.224 eV (b) in p-type InAs as a function
of the magnetic field. The temperature is  $T = 12 \ \mbox{K}$
and the free hole concentration is $p = 10^{19} \ \mbox{cm}^{-3}$.
In (a) the Fermi level is indicated by the dashed line.}
\label{eCR}
\end{figure}

We've examined e-active CR at a photon energy of 0.224 eV in $p$-type
InAs at $T = 12 \ \mbox{K}$ with a free hole density of
$10^{19} \ \mbox{cm}^{-3}$. In Fig.~\ref{eCR} (a) a fan diagram consisting
of the computed $k = 0$ valence band energies is plotted as a function
of the magnetic field. The field-dependent Fermi level is shown as a
nearly horizontal dashed line and the three strongest e-active CR transitions
are indicated by vertical arrows whose relative thicknesses correspond to
the strengths of the three CR absorption transitions. The corresponding
e-active CR absorption spectra is shown in Fig.~\ref{eCR} (b). The most
pronounced e-active transition near 95 Tesla is the sharp interladder
transition $H_{0,2} \rightarrow L_{1,5}$. Another strong interladder
transition $H_{1,3} \rightarrow L_{2,6}$ is seen near 60 Tesla.
This interladder transition has a lot of oscillator strength but is broader
by virtue of the high degree of nonparabolicity in the valence bands.

We now consider the effects of Mn doping on the computed e-active
cyclotron resonance absorption spectra in Fig.~\ref{eCR}. As in
Fig.~\ref{eCR} we will assume that the samples remain paramagnetic
in the presence of Mn doping. In section \ref{valence subband structure},
the Mn doping effect on the valence band structure of In$_{1-x}$Mn$_x$As
was briefly addressed. It was noted that increasing the Mn doping
concentration, $x$, can shift the first HH state $H_{-1,1}$ to the
top of the valence band and shift the LH state $L_{1,5}$ to a lower
position, thus changing the average spin state of the carriers in the
system. The Mn doping has further effects on CR absorption. Doping with
Mn impurities greatly enhances carrier scattering thus increasing
the line width and reducing the strength of the CR absorption.
Manganese doping also shifts the CR peak positions since the $sp-d$
exchange Hamiltonian, $H_{Mn}$ changes the valence subband structure.
Finally, we note that Mn doping is also responsible for the observed
ferromagnetism and this will be discussed in the next section.

\begin{figure}[tbp]
\includegraphics[scale=0.32]{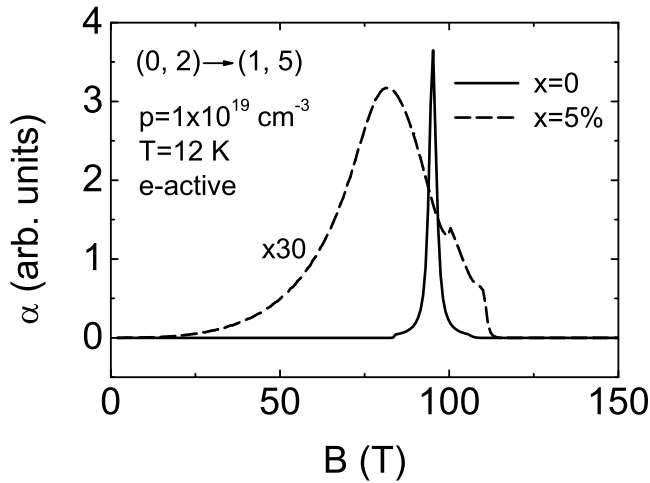}
\caption{The e-active CR transition $H_{0,2} \rightarrow L_{1,5}$ in p-type
In$_{1-x}$Mn$_x$As as a function of the magnetic field for x = 0\%
(solid line) and x=5\% (dashed line). The temperature is $T = 12 \ \mbox{K}$,
the free hole concentration is $p = 10^{19} \ \mbox{cm}^{-3}$, and the sample
is assumed to be paramagnetic.}
\label{MnOnE}
\end{figure}

In Fig.~\ref{MnOnE}, we illustrate the effect of Mn doping on the
strong e-active interladder CR transition $H_{0,2} \rightarrow L_{1,5}$
in Fig.~\ref{eCR}. Two e-active CR absorption
peaks are plotted for two different values of the Mn concentration,
namely $x=0\%$ and $x=5\%$. The strong e-active transition seen in InAs (solid line)
in the absence of Mn doping has been discussed previously and is also show in
Fig.~\ref{eCR}. The e-active CR absorption peak in In$_{0.95}$Mn$_{0.05}$As
(dashed line) is shifted relative to the corresponding InAs feature and reduced
in strength by about a factor of 30. The reduction in strength in going from
x=0\% to x=5\% comes primarily from the Fermi filling effect since the $L_{1,5}$
state contains fewer holes when the sample is doped with Mn ions at a hole density
of $10^{19}\mbox{cm}^{-3}$. We note that absorption takes place not only near
the $\Gamma$ point, but also in regions away from the zone center. Even though
we broaden both CR absorption peaks by the same line width (4 meV), the Mn doped
sample has a much broader line shape due to the change in the
nonparabolic Landau level subband structure brought about by the exchange
interaction.

\begin{figure} [tbp]
\includegraphics[scale=0.5]{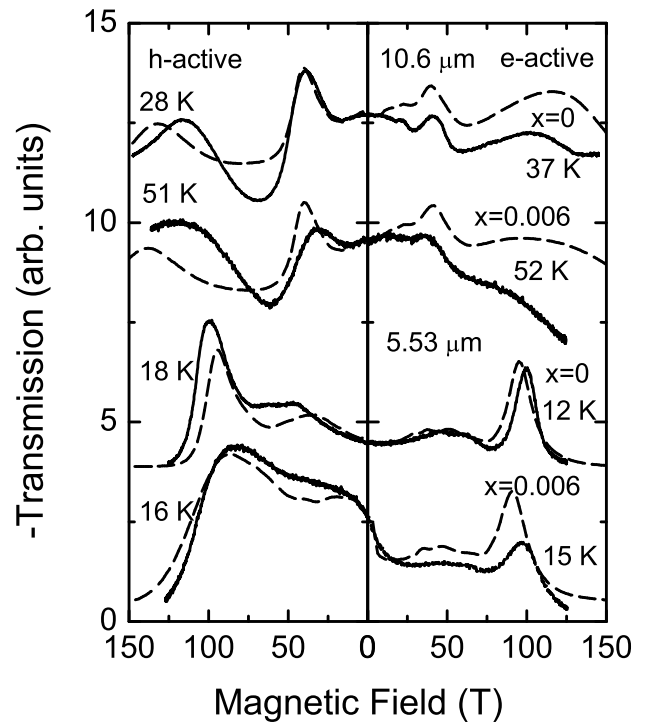}
\caption{Experimental and theoretical CR absorption. Solid lines are
experimental CR spectra as a function of magnetic field for h-active and
e-active polarizations. Corresponding theoretical calculations are shown in
dashed lines.}\label{exp-theo}
\end{figure}

We have performed CR transmission measurements on bulk In$_{1-x}$Mn$_x$As
samples with Mn concentrations x=0\% and x=0.6\% at several temperatures.
In all cases, the samples were found to be in the paramagnetic phase.
The calculated and the experimental CR transmission spectra are shown in
Fig.~\ref{exp-theo} for both e-active (right panel) and h-active (left panel)
polarizations. As seen in Fig.~\ref{exp-theo}, there is good agreement between
theory and experiment. As discussed above, the e-active CR absorption is
determined by the $HH\rightarrow LH$ transitions. The main
contribution to the h-active CR absorption (left panel in
Fig.~\ref{exp-theo}) comes from the transitions within the
heavy hole ladder.

Using both h and e-active CR in p-type dilute magnetic semiconductors,
one may hope to fully characterize the complicated valence band structure
in these important materials.

\subsection{Cyclotron resonance in ferromagnetic thin films}

In addition to studying bulk paramagnetic p-type InMnAs, we also
studied thin ferromagnetic films consisting of InMnAs on (Al,Ga)Sb.
The ferromagnetic films are interesting since they can provide
insight into the interaction of the free carriers and Mn spins. The
interaction of free carriers with localized spins plays an important
role in a variety of physical situations in metals
\cite{Anderson61.41,Kittel68.1,Hewson}. Carriers near a magnetic ion
are spin-polarized, and can mediate an indirect exchange interaction
between magnetic ions.  The hole-induced ferromagnetism realized in
Mn-doped III-V semiconductors
\cite{Munekata91.1011,Munekata93.2929,Ohno96.363} has provided a
novel system in which to study itinerant carriers interacting with
localized spins in the dilute limit. Various mechanisms have been
suggested but the microscopic origin of carrier-induced
ferromagnetism is still controversial
\cite{Dietl00.1019,Konig00.5628,Kaminski02.247202,Zarand02.047201}.
One of the unanswered questions is the nature of the carriers, i.e.,
whether they are in the impurity band ($d$-like), the delocalized
valence bands ($p$-like), or some type of mixed state.

In our studies on the films, we found a clear observation of hole
cyclotron resonance (CR) in ferromagnetic InMnAs/(Al,Ga)Sb. This
demonstrates the existence of at least some delocalized $p$-like
carriers. In addition, this is the first CR study in any ferromagnet
covering temperatures both below and above the Curie temperature
($T_c$) \cite{Goy73.7380,Chin81.7380}. In all samples studied, we
observed two resonances, both of which exhibited an unusual
temperature dependence in their position, intensity, and width. The
lower-field resonance showed an abrupt reduction in width with a
concomitant decrease in the resonance field above $T_c$. The
higher-field resonance, which was absent at room temperature,
suddenly appeared above $T_c$, rapidly grew with decreasing
temperature, and became comparable in strength to the lower-field
resonance at low temperature. We ascribe these resonances to the two
fundamental CR transitions expected for delocalized holes in the
valence band of a Zinc-Blende semiconductor in the magnetic quantum
limit.

In our 8-band {\bf k$\cdot$p} theory ferromagnetism is treated
within a mean-field approximation.  Our results show that the
temperature dependent CR peak shift is a direct measure of the
carrier-Mn exchange interaction.

\begin{table}[bth]
\caption{Characteristics of the In$_{1-x}$Mn$_x$As/Al$_y$Ga$_{1-y}$Sb
samples. Densities and mobilities are room temperature values.
$m_A$ and $m_B$ are the low-temperature cyclotron masses for the two lines
(see Fig. \ref{typical}).}
\label{table1}
\begin{ruledtabular}
\begin{tabular*}{\hsize}{l@{\extracolsep{0ptplus1fil}}c@{\extracolsep{
0ptplus1fil}}c@{\extracolsep{0ptplus1fil}}c@{\extracolsep{0ptplus1fil}}c}
Sample No. & 1 & 2 & 3 & 4\\
\colrule
$T_c$ (K) & 55 & 30 & 40 & 35\\
Mn content $x$ & 0.09 & 0.12 & 0.09 & 0.12\\
Al content $y$ & 0 & 0 & 0 & 1\\
Thickness (nm) & 25 & 9 & 31 & 9\\
Density (cm$^{-3}$) & 1.1$\times$10$^{19}$ & 4.8$\times$10$^{19}$ &
1.1$\times$10$^{19}$ & 4.8$\times$10$^{19}$\\
Mobility (cm$^2$/Vs) & 323 & 371 & 317 & 384\\
$m_A$/$m_0$ & 0.0508 & 0.0525 & 0.0515 & 0.0520\\
$m_B$/$m_0$ & 0.122 & 0.125 & 0.125 & 0.127\\
\end{tabular*}
\end{ruledtabular}
\label{sample characeristics table}
\end{table}

The samples were In$_{1-x}$Mn$_x$As/Al$_y$Ga$_{1-y}$Sb
single heterostructures containing high densities ($\sim$10$^{19}$
cm$^{-3}$) of holes. They were grown by low temperature molecular beam epitaxy
on GaAs (100) substrates \cite{Slupinski02.1326}.
Unlike the $n$- and $p$-type films we studied earlier
\cite{Matsuda01.219,Zudov02.161307,Khodaparast03.107,Sanders03.6897},
the samples in the present work were ferromagnetic.

Table~\ref{sample characeristics table} summarizes the characteristics
of the four In$_{1-x}$Mn$_x$As/Al$_y$Ga$_{1-y}$Sb samples used in our
study. Sample 1 was annealed at 250 $^{\circ}$C after growth, which
increased the Curie temperature, $T_c$, by $\sim$10 K
\cite{Hayashi01.1691,Potashnik01.1495}.
We performed CR measurements using ultrahigh pulsed
magnetic fields generated by the single-turn coil technique
\cite{Zudov02.161307,Nakao85.1018}.
We used circularly-polarized radiation with wavelengths of 10.6 $\mu$m,
10.2 $\mu$m, 9.25 $\mu$m (CO$_2$ laser), and 5.527 $\mu$m (CO laser), and
the transmitted radiation was detected using a
fast HgCdTe detector. A multi-channel digitizer recorded the
signals from the pick-up coil and the detector.

\begin{figure}
\begin{center}
\includegraphics[scale=0.9]{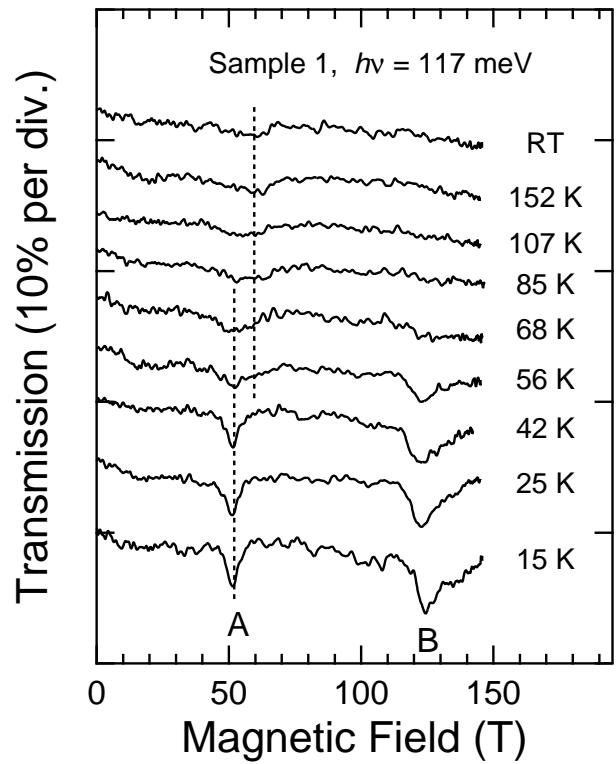}
\caption{CR spectra for sample 1. The transmission of h-active
circularly polarized 10.6 $\mu$m radiation is plotted vs.
magnetic field at different temperatures.}
\label{typical}
\end{center}
\end{figure}

Figure \ref{typical} shows the transmission of the 10.6 $\mu$m
(photon energy $h\nu=117 \ \mbox{meV}$) beam
through sample 1 at various temperatures as a function of
the magnetic field. From room temperature down to slightly above the
Curie temperature, $T_c$, a broad feature (labeled `A') is
observed with almost no change in the line intensity,
position, and width with decreasing temperature.  However,
at $\sim$68 K, which is sill above $T_c$, we observe an abrupt
and dramatic change in the spectra.  As the temperature decreases,
a significant reduction in the linewidth and a sudden shift to a
lower magnetic field are observed along with a rapid increase in
the line intensity.
In addition, a second feature (labeled `B') suddenly appears at
a magnetic field of $\sim$ 125 Tesla, which also rapidly grows
in intensity with decreasing temperature and saturates at low
temperature in a manner similar to that observed in feature A. Note that
{\em the temperature at which these unusual sudden CR changes occur
($T_c^*$) is higher than $T_c$.}

\begin{figure}
\begin{center}
\includegraphics[scale=1.0]{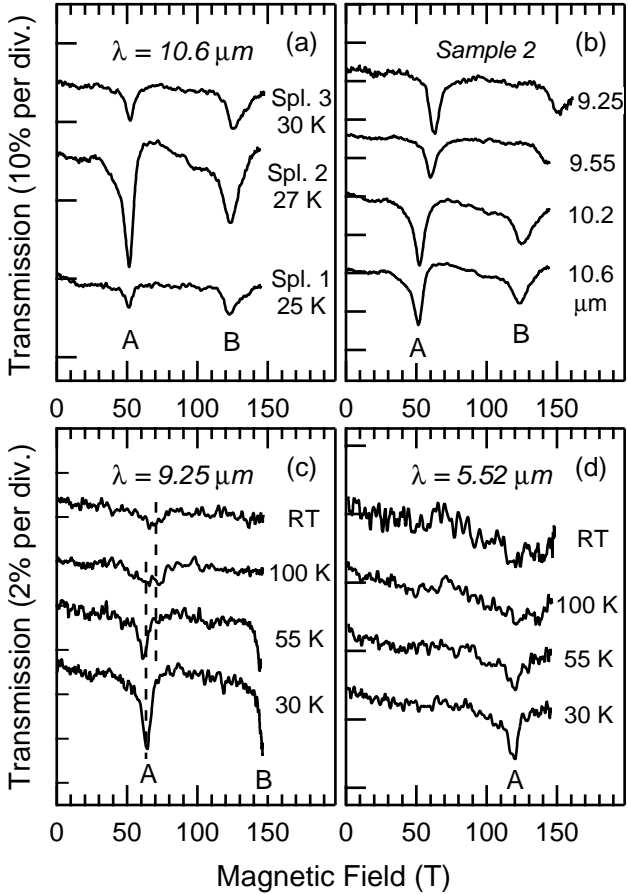}
\caption{
(a) Low temperature CR spectra for three samples at
10.6 $\mu$m.  (b) Wavelength dependence of the CR spectra for
sample 2 at 27 K.  CR spectra for sample 1
at different temperatures at (c) 9.25 $\mu$m and (d) 5.52 $\mu$m.
}
\label{various}
\end{center}
\end{figure}

The observed unusual temperature dependence is neither specific to the
particular wavelength ($\lambda$) used or the sample measured.
We observed essentially the same temperature dependent behavior in all the
samples studied.  Figure \ref{various} (a) shows low temperature
CR traces for three samples at 10.6 $\mu$m. Both features A and B are clearly
observed but their intensities and linewidths vary from sample
to sample. Figure \ref{various} (b) displays the wavelength dependence of
the CR spectra for sample 2.  We can see that both lines shift
to higher magnetic fields with decreasing wavelength (i.e.,
increasing photon energy), as expected.
Figures \ref{various} (c) and \ref{various} (d) show data at different
temperatures for sample 1 measured at wavelength of 9.25 $\mu$m and
5.52 $\mu$m, respectively. The temperature dependence observed at these
shorter wavelengths is similar to what was observed at 10.6 $\mu$m. The
observations of CR with essentially the same masses in samples with
different buffer layers (GaSb or AlSb) exclude the possibility that we
are observing hole CR in the buffer.  We also confirmed
the absence of CR in a control sample which consisted of only a GaSb layer
grown on GaAs.  All these facts confirm the universality of the effects we
observed and their relevance to ferromagnetic order.

The clear observation of CR indicates that {\em at least}
a fraction of the holes are delocalized. This is in agreement with our
measurements on low-$T_c$ films
\cite{Khodaparast03.107,Sanders03.6897} which shows
that the two resonance spectra are similar although the resonances
were much broader and the temperature dependence was much weaker.
However, extensive earlier attempts to observe CR in GaMnAs
\cite{Matsuda97.42} did not detect any sign of resonant absorption
within the magnetic field and wavelength ranges in which
both light hole (LH) and heavy hole (HH) CR in GaAs were expected.
This fact may indicate that the holes in GaMnAs are strongly localized,
that the mixing of $p$- and $d$-like states makes the effective masses of
holes extremely large, or that scattering is too strong for
the cyclotron resonance condition, $\omega_c\tau \gg 1$, to hold.
In any case, it appears that the carriers mediating the Mn-Mn exchange interaction
are considerably more localized in GnMnAs than in InMnAs, consistent with
recent optical conductivity \cite{Hirakawa02.193312}
and photoemission experiments \cite{Okabayashi02.161203}.

Feature A becomes strikingly narrow at low temperatures. The estimated CR mobility is
4$-$5 $\times$ 10$^3$ cm$^2$/Vs, which is one order of magnitude larger than
the low temperature mobilities measured by the Hall effect
(see Table~\ref{sample characeristics table}). We speculate that
this is associated with the suppression of localized spin
fluctuations at low temperature. A similar effect has been observed
in (II,Mn)VI systems \cite{Brazis02.73}.
Spin fluctuations become important when a band carrier
simultaneously interacts with a limited number of localized spins. This occurs,
for example, for magnetic polarons and for electrons in (II,Mn)IV quantum
dots. The strong in-plane localization by the magnetic field may also
result in a reduction of the number of spins which a band carrier
feels, thus increasing the role of spin fluctuations.

It is important to emphasize that the temperature at which significant spectral changes
start to appear ($T_c^*$) is consistently higher than $T_c$ in all the samples.
This fact may be explainable in light of a recent Monte Carlo study
\cite{Schliemann01.165201}
which suggested that {\em short-range} magnetic order and finite {\em local}
carrier spin polarization are present for temperatures substantially higher than $T_c$.
A more recent theoretical study \cite{Alvarez02.277202} explicitly predicts the existence
of $T_c^*$, corresponding to {\em clustering}. Any such local order should
result in modifications in band structure, which in turn modify CR spectra.

\begin{figure}[b]
\begin{center}\leavevmode
\includegraphics[scale=0.6]{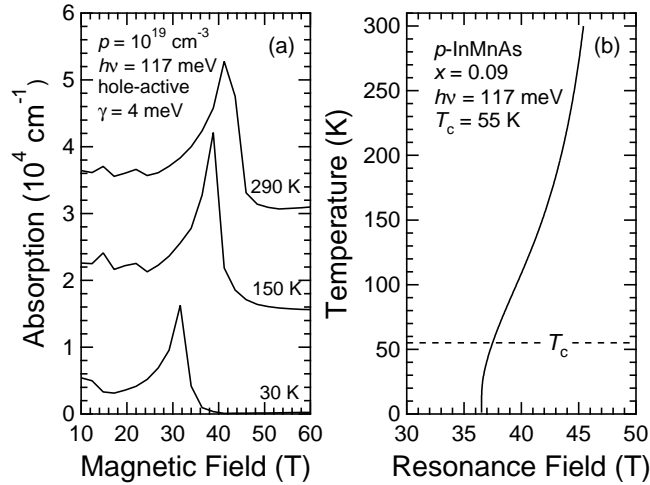}
\caption{(a) Theoretical CR spectra for sample 1 showing a shift of peak A
with temperature.  (b) Calculated temperature-dependence of the
resonance field for peak A in sample 1.
}
\label{theory}
\end{center}
\end{figure}

In our theoretical model,
peak A can be identified as the H$_{-1,1}$ $\rightarrow$ H$_{0,2}$
transition\cite{Sanders03.6897,Sanders03.449}.
We attribute the temperature dependent peak shift to the increase
of carrier-Mn exchange interaction resulting from the increase
of magnetic ordering at low temperature. The carrier-Mn exchange
interaction interaction is governed by the Hamiltonain $H_{Mn}$ in
Eq.~\ref{HTotal} and depends explicitly on the expectation value
of the Mn spin, $\langle S_z \rangle$, in the mean field approximation.
Our calculated CR spectra are shown in
Fig.~\ref{theory} (a) for bulk In$_{0.91}$Mn$_{0.09}$As with only a
minimal broadening of 4 meV.  The figure shows the shift of peak A with
decreasing temperature.  Note that the
peak in a bulk system occurs at room temperature at a magnetic field
of $\sim$40 T as opposed to the heterostructures where the resonance occurs
at a magnetic field of $\sim$50 T due to quantum confinement and strain effects.

It is easy to obtain an
exact analytical expression for this shift since it involves only the lowest
two manifolds in our model ($n = -1$, which is 1 dimensional, and $n = 0$,
which factors into two 2$\times$2 matrices for $k_z = 0$).
Furthermore, to simplify the final expressions, we neglect the small terms
arising from the interaction with remote bands. With these simplifications,
the cyclotron energy (at the center of the Landau subbands) has the form:
\begin{eqnarray}
E_{CR} &=& -\frac{E_g}2 + \frac14 x \langle S_z \rangle
(\alpha-\beta)  \nonumber \\
& &
+ \sqrt{\left[\frac{E_g}2 -\frac14 x \langle S_z \rangle
(\alpha-\beta)\right]^2+E_p \mu_B B} ,
\label{dE}
\end{eqnarray}
where $E_g$ is the energy gap, $E_p$ is related to the Kane momentum matrix
element $P$ by $E_p=\frac{\hbar^2 P^2}{2 m_0}$, $\alpha$ and $\beta$ are $s$-$d$
and $p$-$d$ exchange constants, and $x \langle S_z \rangle$ is the
magnetization per unit cell.

In the field range of interest ($\sim 40$ T), $\sqrt{E_p \mu_B B}$ is
of the same order as $\frac{E_g}2$, while the exchange interaction is much
smaller even in the saturation limit. Expanding the square root in Eq.~(\ref{dE}),
we obtain the final expression
\begin{equation}\label{dE1} E_{CR} \approx
\frac{E_g}2\left(\frac1{\delta}-1\right)+ \frac14 x \langle
S_z \rangle (\alpha-\beta)(1-\delta),
\end{equation}
where $\delta = E_g / (E_g^2+4 E_p \mu_B B)^{1/2}$.

If we assume that $E_g$ and $E_p$ do not change appreciably with temperature,
it follows from Eq.~(\ref{dE1}) that the peak shift should follow
the temperature dependence of $\langle S_z \rangle$. This shift directly
measures the carrier-Mn exchange interaction. To obtain quantitative
agreement with the experiment, one should calculate $\langle S_z \rangle$
by taking into account the possibility of short-range ordering, as discussed above
\cite{Schliemann01.165201,Alvarez02.277202}. This effect could substantially modify
the band structure at low magnetic fields. At high magnetic fields, however,
this effect should be smoothed out by the field-induced magnetic ordering.
In the following we neglect this effect and calculate $\langle S_z \rangle$
via standard mean-field theory \cite{Ashcroft}, solving the transcendental equation
\begin{equation}
\label{Sz}
\langle S_z \rangle
= S B_S \left(\frac{g S}{kT}\left[\mu_B B-\frac{3 k T_c \langle S_z
\rangle}{gS(S+1)}\right]\right),
\end{equation}
where $g$ is the free electron $g$ factor, $B_S$ is the Brillouin
function, and $S=\frac52$ is the spin of the magnetic ion.

The temperature dependence of the resonance field, calculated using
Eqs.~(\ref{dE1})-(\ref{Sz}), is presented in Fig. \ref{theory} (b).
Parameters used in the calculation are $x$ = 0.09,
$T_c$ = 55 K, $E_g$ = 0.4 eV, $E_p$ = 21 eV, and $\alpha-\beta=1.5$ eV.
It shows that from room temperature to 30 K the the resonance magnetic
field decreases by $\sim$20\%, approximately the result observed in
the experiment. In addition, we find that the shift is nonlinear with
temperature and the main shift occurs at temperatures well above $T_c$,
which is also consistent with the experiment.

One word of caution should be noted.  While our theoretical
calculations (eq. \ref{dE1}) predict a definite shift in the hole
cyclotron resonance peaks with an increase in the magnetization of
the Mn spins, in apparent agreement with the experimental results,
this agreement might be fortuitous.  There are two possible sources
of discrepancy between theory and experiment that must first be
resolved before a definitive conclusion can be reached. One is that
there could be a possible difference between the Mn concentration
and the {\it effective} Mn concentration, i.e. not all Mn ions
contribute to $\langle S_z \rangle$.  The second is that one must
rule out any contributions from the interface or the (Al,Ga)Sb
layers.

\section{Conclusions}
\label{conclusions}

We have presented a theoretical and experimental study of the
electronic and magneto-optical properties of $p$-type paramagnetic
In$_{1-x}$Mn$_{x}$As dilute magnetic semiconductor alloys and
ferromagnetic p-type In$_{1-x}$Mn$_{x}$As/(Al,Ga)Sb thin films in
ultrahigh magnetic fields oriented along [001]. We used an 8-band
Pidgeon-Brown model generalized to include the wave vector
dependence of the electronic states along $k_{z}$ as well as the
$s-d$ and $p-d$ exchange interactions with the localized Mn $d$
electrons. The Curie temperature is taken as an input parameter and
the average Mn spin is treated in mean field theory.

In the case of p-type InMnAs we find that: (i) Mn doping changes the
spin states of the DMS system. We find that Mn doping shifts the HH
spin down state $H_{-1,1}$ to the valence band edge which, in the
absence of doping, is a LH spin up state $L_{1,5}$. The spins of the
itinerant holes are thereby flipped through the exchange interaction
with the localized Mn moments. (ii) Two primary CR absorption peaks
are present for magnetic fields up to 300 Tesla for h-active
polarization with a photon energy of $0.117$ eV. These peaks
correspond to $HH\rightarrow HH$ and $LH\rightarrow LH$ transitions.
(iii) CR transitions take place at not only the zone center, but
also in the regions away from the $\Gamma$ point. The camel back
structures in some of the Landau subbands enhance this effect. (iv)
E-active CR absorption in p-type InMnAs is observed and arises due
to the complexity of the valence bands. These transitions are seen
to arise from transitions between \emph{different} hole ladders
($HH\rightarrow LH$ or $LH\rightarrow HH$). Because h-active CR
takes place in the same ladder, both e-active CR and h-active CR can
be used to explore the complex valence bands.

In calculating the valence band structure and CR absorption, we find
that valence band structure and CR absorption strongly depends on
the material parameters. The LH transitions is more affected by
changes in the values of these parameters than the HH transitions.
This in turn can be used as a method to evaluate these parameters.
The CR spectra strongly depends on hole densities through the
dependence on the Fermi energy in the valence bands. This can also
be used to estimate the hole densities, which becomes more
meaningful due to the difficulty in traditional methods to measure
the carrier densities because of the anomalous Hall effect.

In ferromagnetic p-InMnAs/(Al,Ga)Sb, two strong CR peaks are
observed which shift with position and increase in strength as the
Curie temperature is approached from above.  The shift in the peaks
is predicted theoretically and an expression for the shift is given.
This transition takes place well above the Curie temperature and can
be attributed to the increase in magnetic ordering at low
temperatures.

\appendix

\section{Validity of 8 band $ \vec{k}\cdot \vec{P}$  calculations in high magnetic fields}

\begin{figure} [tbp]
\includegraphics[scale=0.5]{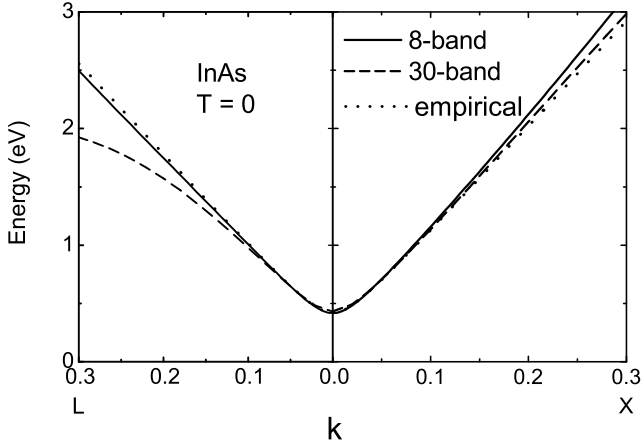}
\caption{The conduction band electronic structure in the X and L
directions for different calculations.  The solid line is based on 8
band $ \vec{k}\cdot \vec{P}$, the dashed line is for full-zone, 30
band $ \vec{k}\cdot \vec{P}$ theory and the dotted line is based on
the empirical nonparabolic formula given by eq. \ref{app3}. As can
be seen, the 8 band model accurately models the full zone structure
up to about 20\% of the zone boundary.}
\label{ConductionXL}
\end{figure}

An important question which arises is ''How accurate is an 8 band $
\vec{k}\cdot \vec{P}$ model for magnetic fields that are 100 T or
more?''  One can make a simple estimate for the accuracy of the 8 band $
\vec{k}\cdot \vec{P}$  model. The presence of a magnetic field in
the $\widehat{z}$ direction quantizes the carrier orbits in the
$x-y$ plane.  These quantized orbits are approximately given by
\begin{equation}
\label{app1}
\frac{\hbar^{2}}{2m^{*}}(k_{x}^{2}+k_{y}^{2})=(n+1/2)
\hbar\omega_{c}
\end{equation}
where $\omega_{c}=eB/m^{*}c$ is the cyclotron frequency.
Factoring out the effective mass we obtain
\begin{equation}
\label{app2} \frac{\hbar^{2}}{2}(k_{x}^{2}+k_{y}^{2})=(n+1/2)
\frac{\hbar e B}{c}.
\end{equation}
Note that this is essentially equivalent to the Bohr-Sommerfeld
quantization conditions for circular orbits as described in
Kittel.\cite{Kittel1}

The maximum value for $k_x$ on an orbit occurs when $k_y=0$.  When
this occurs, the magnetic field is related to $k_x$ by:
\begin{equation}
\label{app3} B = \frac{ \hbar k_x^2 c}{(2n+1) e}.
\end{equation}
While this is a simplified model (the hole orbits are not circular
but have a complex shape due to warping of the bands), it allows us
to estimate for what magnetic fields for which the 8 band
$\vec{k}\cdot \vec{P}$ theory is valid.

In Fig.~\ref{ConductionXL} we plot the conduction bands as
calculated by both 8 band (solid line) and {\it full-zone} 30 band
(dashed line) $ \vec{k}\cdot \vec{P}$
calculations.\cite{Cardona66.530,Bailey90.3423, Stanton90.231} Also
plotted is the empirical expression for a nonparabolic conduction
band (dotted line)
\begin{equation}
\label{app4} \varepsilon (1 + \alpha \varepsilon) =
\frac{\hbar^{2}k^2}{2m^{*}}
\end{equation}
where $\alpha$ is the nonparabolicity parameter.  As can be seen,
the agreement between the 30 band full-zone calculation and the
8-band model is very good up to about 15-20\% of the X and L points
at the brillouin zone boundary.

\begin{figure} [tbp]
\includegraphics[scale=0.5]{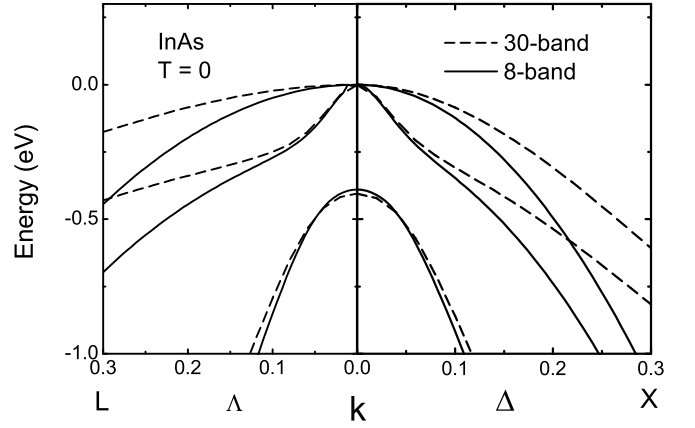}
\caption{Heavy, light and spin-split valence bands for InAs out to
30\% of the zone boundary.  The 8 band model (solid line) is
compared to the full-zone model (dashed line) and shows agreement
out to 10-15\% of the zone boundary.}
\label{val20}
\end{figure}

In Fig.~\ref{val20}, we see the three valences bands plotted for
both the 8 band model (solid lines) and the full-zone 30 band model
(dashed lines).  Here the agreement is not as good as the conduction
band with the 8 band model agreeing with the full-zone model only to
about 10-15\% of the zone boundary. (Note that we have not tried to
optimize the parameters of the 30 band model to produce maximum
agreement).

Since the cyclotron resonance transitions involve the $n=1
\rightarrow n=0$ transition, we would expect that the 8 band model
would be valid up to magnetic fields given by Eq.~\ref{app3} with
n=1 and $k_x = $ 0.1 to 0.15 $k_{zb}$ where $k_{zb} = 2\pi/a$ is the
zone boundary wavevector in the $\widehat{x}$ direction and $a$ is
the lattice constant.  For InAs, the lattice constant is
6.06${\AA}$.

Substituting into Eq.~\ref{app3}, we see that if the 8 band
model gives agreement with the full zone for 10\% of the zone
boundary, then the maximum magnetic field is about 234 T.  For 15\%,
the maximum field is 520 T, and for 20\% the field is 930 T.

These are rather large fields so it is not surprising that good
agreement is reached for both the heavy hole transitions (at $ \sim
$ 50 T) and the light hole transitions (at $\sim $ 150 T).  The
higher order transitions occurring at around 400 T in the
experimental data do not agree as well with the theoretical
calculations and might be attributed to the breakdown in the 8 band
model.

\begin{acknowledgments}
This work was supported by the National Science Foundation ITR
program through grant DMR-032547 and by DARPA MDA 972-00-1-0034.
\end{acknowledgments}

\bibliography{paper}

\end {document}